\documentclass[aps,pre,showpacs,twocolumn,superscriptaddress,showkeys,reprint]{revtex4-1}
\usepackage{graphicx}
\usepackage{color}

\begin{document}
\title{Experimental Observation of the Spectral Gap in Microwave $n$-Disk Systems}
\author{S.~Barkhofen}
\affiliation{Fachbereich Physik, Philipps-Universit\"{a}t Marburg, Renthof 5,
35032 Marburg, Germany}
\author{T.~Weich}
\affiliation{Fachbereich Physik, Philipps-Universit\"{a}t Marburg, Renthof 5,
35032 Marburg, Germany}
\affiliation{Fachbereich Mathematik, Philipps-Universit\"{a}t Marburg, Hans-Meerwein-Stra{\ss}e,
35032 Marburg, Germany}
\author{A.~Potzuweit}
\affiliation{Fachbereich Physik, Philipps-Universit\"{a}t Marburg, Renthof 5,
35032 Marburg, Germany}
\author{H.-J.~St\"{o}ckmann}
\affiliation{Fachbereich Physik, Philipps-Universit\"{a}t Marburg, Renthof 5,
35032 Marburg, Germany}
\author{U.~Kuhl}
\affiliation{Laboratoire de Physique de la Mati\`{e}re Condens\'{e}e, CNRS UMR 7336, Universit\'{e} de Nice
Sophia-Antipolis, F-06108 Nice, France}
\affiliation{Fachbereich Physik, Philipps-Universit\"{a}t Marburg, Renthof 5,
35032 Marburg, Germany}
\author{M.~Zworski}
\affiliation{Department of Mathematics, University of California, Berkeley, California 94720, USA}

\date{\today}
\begin{abstract}
Symmetry reduced three-disk and five-disk systems are studied in a microwave setup. Using harmonic inversion the distribution of the imaginary parts of the resonances is determined. With increasing opening of the systems, a spectral gap is observed for thick as well as for thin repellers and for the latter case it is compared with the known topological pressure bounds. The maxima of the distributions are found to coincide for a large range of the distance to radius parameter with half of the classical escape rate. This confirms theoretical predictions based on rigorous mathematical analysis for the spectral gap and on numerical experiments for the maxima of the distributions.
\end{abstract}


\maketitle

In semiclassical physics we investigate asymptotic quantum-to-classical correspondence when an effective Planck constant is small. Examples for closed systems are the Weyl law \cite{wey12c} which gives densities of quantum states using classical phase space volumes and the Gutzwiller trace formula\cite{gut71,gut90} which describes the fluctuations of these densities in terms of classical periodic orbits and their stability \cite{gut90}.

For open systems the correspondence between classical and quantum quantities \cite{gra10, non11} is more delicate as energy shells are noncompact and real eigenvalues of the Hamiltonian become complex resonances \cite{moi98,zwo99a,kuh13}. The imaginary parts of resonances are always negative and they correspond to the rate of decay of unstable states.

For open chaotic systems the Weyl law is replaced by its fractal analogue which gives asymptotics of the number of resonances with bounded imaginary parts in terms of the dimension of the fractal repeller (see Refs.~\cite{sjo90,sjo07} for mathematical studies, Refs.~\cite{sch00d,lin02b,lu03,sch09} for numerical studies, and Refs.~\cite{pot12} for recent experimental work). Studying the distribution of the imaginary parts of resonances \cite{lu03,sch09} does not have a closed system analogue.

\begin{figure}[b]
\includegraphics[width=0.8\columnwidth]{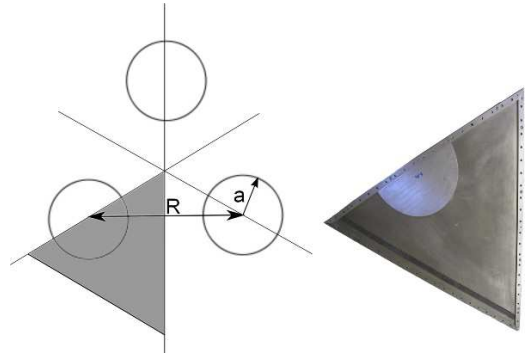}
\caption{\label{fig:Foto} (color online)
A sketch of a three-disk system is shown on the left side, where one fundamental domain is shaded. On the right side a photograph of the experimental cavity without top plate supporting the disk inset and the absorber is presented.}
\end{figure}

A paradigm for systems with fractal repellers is the $n$-disk scattering system (see Fig.~\ref{fig:Foto}). It was introduced in the 1980s by Ikawa in mathematics \cite{ika88} and by Gaspard and Rice \cite{gas89a,gas89b,gas89d} and Cvitanovi{\'c} and Eckhardt \cite{cvi89} in physics. It is given by $n$ hard disks with centers forming a regular polygon. The distance between the centers is denoted by $R$ and the disk radius by $a$; $ R/a$ determines the system up to scaling (see Fig.~\ref{fig:Foto}).

The quantum system is described by the Helmholtz equation
\begin{equation}
-\nabla^2\psi_n=k_n^2\psi_n, \ \ \text{$ \psi_n = 0 $ on disc boundaries.}
\end{equation}
The quantum resonances $k_n = \mathrm{Re}\, k_n + i \mathrm{Im}\, k_n$, are the complex poles of the scattering matrix. For the three-disk system this scattering matrix is expressed using Bessel functions and that allowed Gaspard and Rice \cite{gas89d} to calculate the quantum resonances numerically.

Classically, particle trajectories are given by straight lines reflected by the disks. From periodic trajectories a wide range of classical quantities such as the classical escape rate, the fractal dimension of the repeller and the topological pressure can be calculated using the Ruelle zeta function \cite{gas89b},
\begin{equation}
\label{eq:zeta}
 \zeta_\beta(z)= \prod\limits_p \left[1-\frac{\exp(-zT_p)}{\Lambda_p^\beta} \right]^{-1}
\end{equation}
where the product runs over the primitive periodic orbits, $T_p$ are the corresponding period lengths, and $\Lambda_p$ are the stabilities. The topological pressure $P(\beta)$ is then defined as the largest real pole of $\zeta_\beta(z)$. An effective method for its calculation is the cycle expansion \cite{cvi89, eck95b}. The classical escape rate is given by $\gamma_\mathrm{cl}=-P(1)$ and the reduced Hausdorff dimension $d_H$ of the fractal repeller by the Bowen pressure formula $P(d_H) = 0$ \cite{bow79}.

\begin{figure}[t]
  \mbox{\parbox{0.0cm}{\raisebox{4cm}[0pt][0pt]{(a)}}\hspace*{-.75cm}
  \includegraphics[height=5cm]{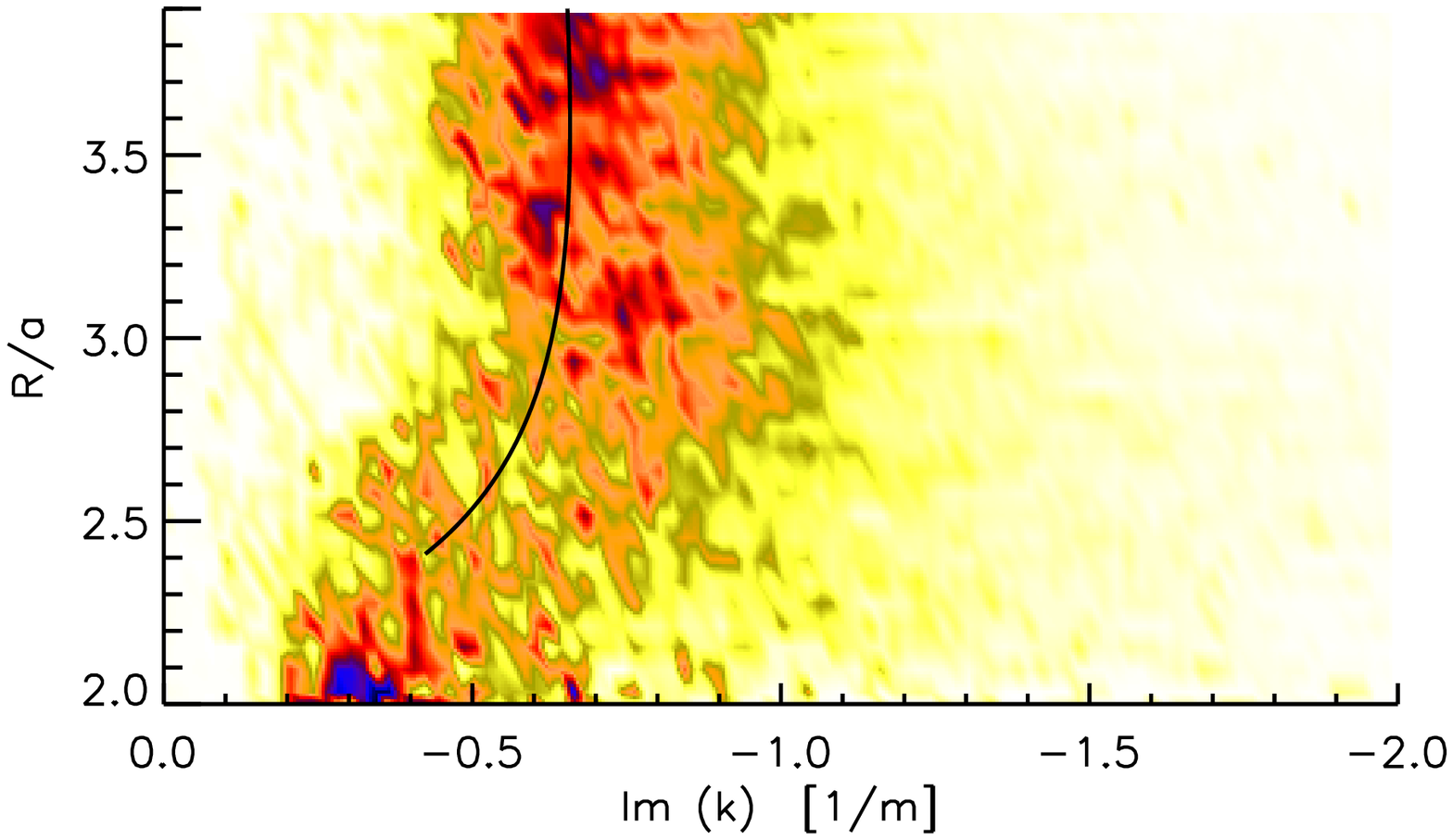}}\\[-.3cm]
  \mbox{\parbox{0.0cm}{\raisebox{4cm}[0pt][0pt]{(b)}}\hspace*{-.75cm}
  \includegraphics[height=5cm]{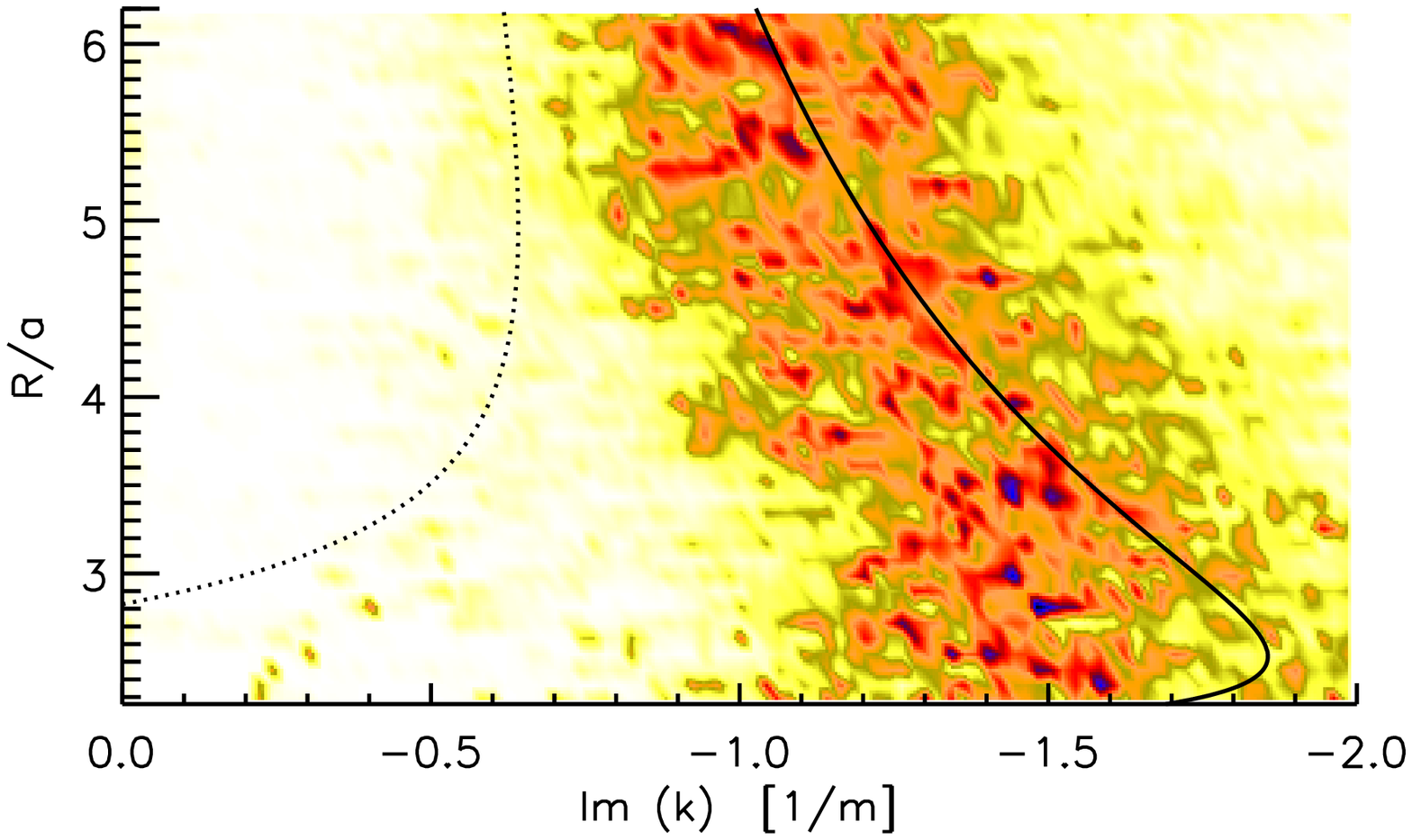}}\\[-.3cm]
  \mbox{\parbox{0.0cm}{\raisebox{4cm}[0pt][0pt]{(c)}}\hspace*{-.75cm}
  \includegraphics[height=5cm]{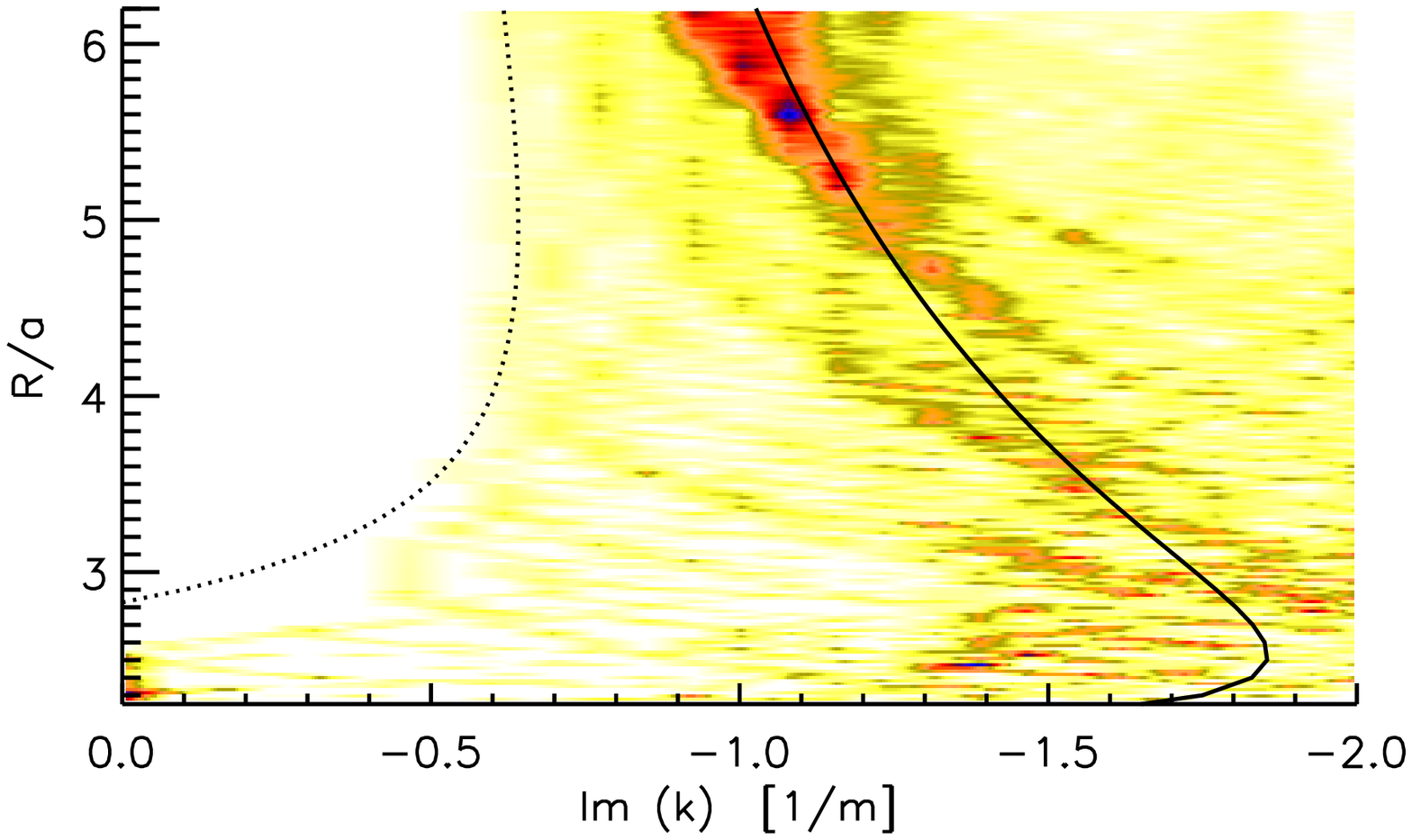}}
  \caption{\label{fig:EscRate} (color online)
  Shade plots of the distribution of $\mathrm{Im}\, k$ using a color code from white to dark blue for (a) the five-disk experiment, (b) the three-disk experiment, and (c) the three-disk simulation as a function of the $R/a$ parameter. The classical escape rate (precisely: $-\gamma_\mathrm{cl}/2$) is the solid black line. $P(1/2)$ corresponds to the dotted line. For the five-disk case the $P(1/2)$ is still positive and hence does not arise in the plot.}
\end{figure}

Ikawa \cite{ika88} and Gaspard and Rice \cite{gas89b} independently described the quantum mechanical spectral gap as a topological pressure, a purely classical quantity. The spectral gap in this context is the smallest $ C $ such that $\mathrm{Im}\, k_n \leq C$. Gaspard and Rice used Gutzwiller's trace formula and semiclassical zeta functions to conclude that $\mathrm{Im}\, k_n \leq P(1/2)$. They confirmed this estimate numerically \cite{gas89a}; this estimate was later proved for general semiclassical systems \cite{non09}. However, this lower bound is not optimal as it does not take into account phase cancellation \cite{non11,pet10} and for weakly open systems this bound is void: for systems with $d_H > 1/2$, $P(1/2) > 0 $. Hence we distinguish between ``thick'' repellers with $d_H \geq 1/2$ [Fig.~\ref{fig:EscRate}(a)] and ``thin'' repellers, $ d_H < 1/2 $ [Fig.~\ref{fig:EscRate}(b)] (see Ref.~\cite{non11}).

The same estimate on the spectral gap was obtained earlier for hyperbolic quotients $\Gamma \backslash {\mathbf H}^2$ \cite{pat76,sul79}, another mathematical model for chaotic scattering \cite{bor07}. There, quantum resonances (poles of the scattering matrix of the surface) are the zeros of the Selberg zeta function and the topological pressure can be calculated explicitly using $\delta$, the dimension of the limit set of $\Gamma$: $P ( \beta ) = \delta - \beta$. The estimate $\mathrm{Im}\, k_n \leq \delta -1/2$ is known to be sharp as $i (\delta -1/2)$ (a bound state when $\delta -1/2 >0$) is a resonance. There are no other resonances for $\mathrm{Im}\,k < P ( 1/2 ) - \epsilon$, for some small $\epsilon > 0$ \cite{nau05}. The question of further improvements for the spectral gap is an active field of mathematical research with deep applications to number theory \cite{non11,bou11}.

An interesting property of the $\mathrm{Im}\, k$ distribution has been observed numerically in Ref.~\cite{lu03}: the imaginary parts of resonances concentrate at $\mathrm{Im}\, k = - \gamma_\mathrm{cl}/2 = P ( 1) /2$, half of the classical escape rate. Although no mathematical result supports this $-\gamma_\mathrm{cl}/2$, the density of resonances for $\mathrm{Im}\,k > - \gamma_\mathrm{cl}/2$ is lower \cite{nau13} than the prediction from the fractal Weyl law \cite{gui04,zwo99b,arXdat12}.

Another connection between the classical escape rate and the quantum spectrum was observed in microwave $n$-disk experiments \cite{lu99,lu00}: the decay of the wave-vector autocorrelation function for small wave vectors is related to the classical escape rate.

In this Letter we focus on the distribution of imaginary parts of resonances and compare spectral gaps and density peaks of the experimental $\mathrm{Im}\,k$ distribution with the topological pressures and the classical escape rates.

The $n$-disk system is simplified by exploiting its $D_n$ symmetry. In this reduction, the two enclosing symmetry axes are hard walls acting as ``mirrors'' (see the shaded area in Fig.~\ref{fig:Foto}). For the quantum mechanical system 0-boundary conditions at the symmetry axes imply that the corresponding scattering resonances are in the $A_2$ representation \cite{eck95b}. The reduced three- and five-disk system is realized using a microwave cavity. The triangular resonator (Fig.~\ref{fig:Foto}) has two metallic side walls of length $1$\,m meeting at $60^\circ$ for the three-disk, and at $36^\circ$ for the five-disk system. Absorbers on the third side model an open end. The ratio $R/a$ is changed by moving a half-disk inset of radius $a=19.5$\,cm along the side wall in steps of 10\,mm. For the three-disk system the range $2.26 \leq R/a \leq 6.17$ was technically accessible; for the five-disk case we had $2 \leq R/a \leq 3.9$. A 0.7\,mm wire antenna was inserted through a hole in the top plate. The height of the cavity $h=6$\,mm leads to a cutoff frequency of 25\,GHz. From 2 to 24\,GHz only the TM$_0$ mode can propagate and the cavity is effectively two dimensional. Hence the equivalence between wave mechanics and quantum mechanics, i.\,e.,\ between the time independent Helmholtz and Schr\"{o}dinger equation, is valid (for more on the setup see Ref.~\cite{pot12}, and for an introduction to microwave billiards, see Chap.~2.2 of Ref.~\cite{stoe99}).

\begin{figure}
\includegraphics[width=.8\columnwidth]{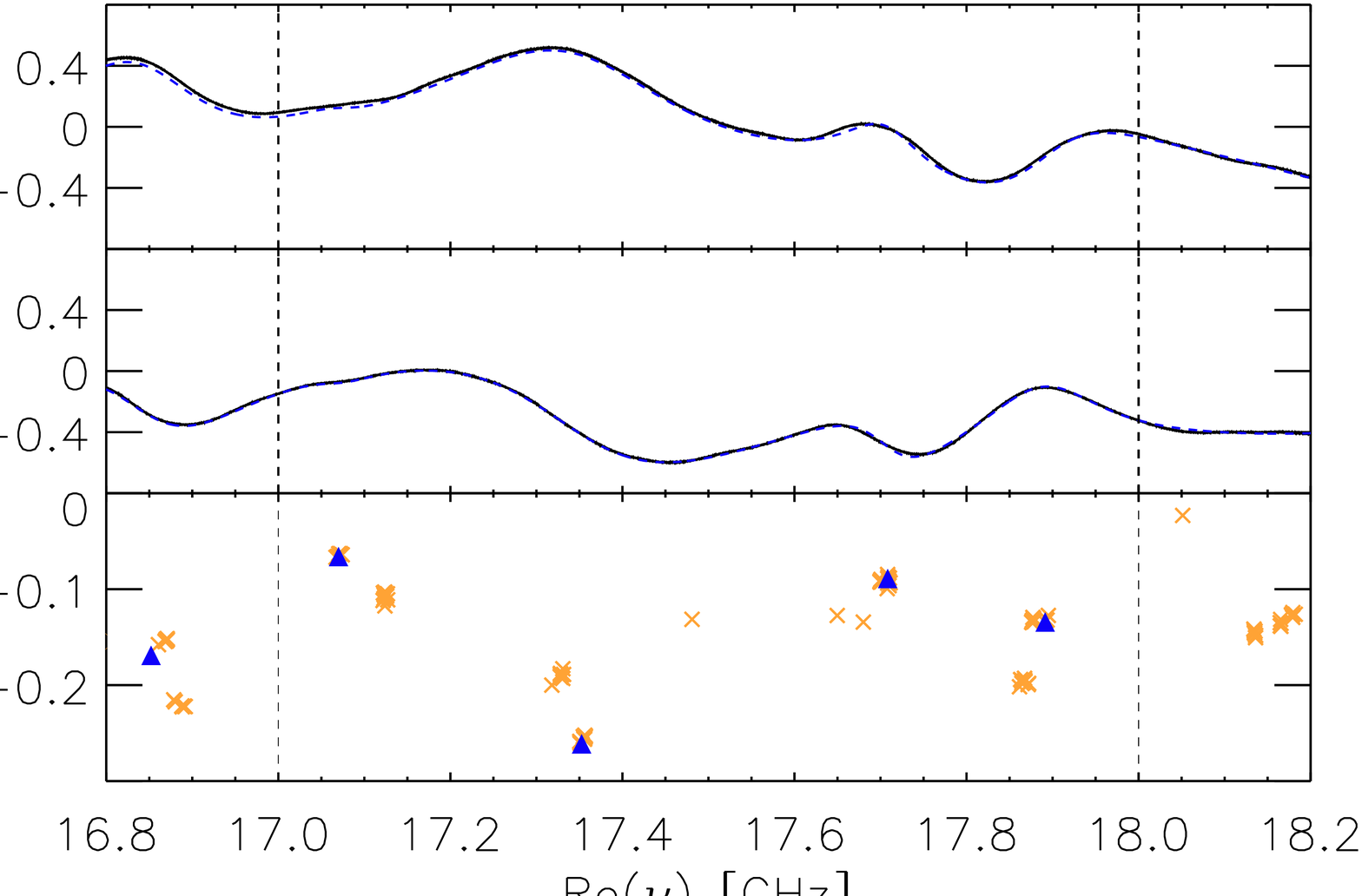}
\caption{\label{fig:resonanceClouds} (color online)
Three-disk system at $R/a = 2.88$: in the lowest panel resonances belonging to good (orange crosses) and the best reconstruction (blue triangles) are shown in the complex plane within a small frequency range. The upper two panels
show the real and imaginary parts of the measured signal (black, dashed), and of the best reconstruction (blue) in this window (vertical lines) based on the poles marked by the blue triangles. }
\end{figure}

\begin{figure*}
\includegraphics[width=1.85\columnwidth]{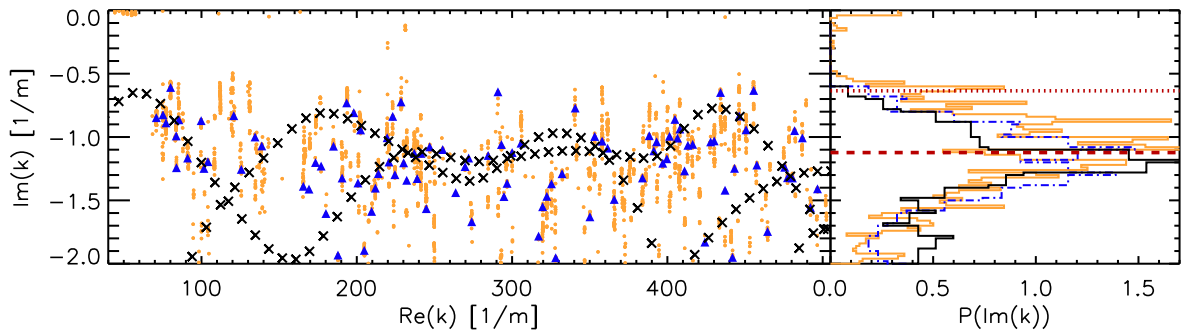}
\caption{\label{fig:WidthDistr} (color online)
In the left panel the resonances for $R/a = 5.5$ in the complex $k$ plane are shown as well as the distribution of the imaginary parts of $k$ in the right panel. The shown $k$ range corresponds to a frequency range from 2 to 24\,GHz. The orange clouds correspond to all resonances resulting in a good reconstruction as well as the orange histogram. Note that many orange dots might overlay each other. Blue triangles and blue dashed-dotted histogram describe the set belonging to the best reconstruction. Black crosses are the numerically calculated poles and the solid black histogram the corresponding distribution. The dotted red line in the right panel is $P(1/2)$, the red dashed line $-\gamma_\mathrm{cl}/2 = P(1)/2$.}
\end{figure*}

Measurements by a vector network analyzer reveal the complex $S$ matrix. Assuming a point-like antenna, the measured reflection signal equals \cite{ste95}
\begin{equation}\label{eq:SumSquaredLorentzians}
  S_{11}(\nu) =1+ \sum\limits_j \frac{ A_j}{\nu^2 - \nu_j^2}
\end{equation}
where $\nu_j$ are the complex valued resonance positions. Extracting $\nu_j$ and ${ {A_j}}$ from the signal is the object of our data analysis: for closed systems and low frequencies the resonances are well separated and a multi-Lorentz fit works. For open systems, where the resonances overlap strongly, that fit does not converge. Therefore we applied the harmonic inversion (HI) on the signal \cite{mai99,kuh08b,pot12}, a sophisticated nonlinear algorithm, to extract the
$ \nu_j$'s from the measured signal. First $S_{11}(\nu)$ is converted into a time signal and discretized, yielding a sequence $c_1,\dots,c_{2N}$. Using the relations between the $c_n$, a matrix of rank $N$ is created. The eigenvalues and eigenvectors of this matrix contain the information about the resonances $\nu_j$ and their residues $A_j$ (see Ref.~\cite{kuh08b}). The procedure yields $N$ resonances; hence $N$ has to be chosen larger than its expected number. Criteria for eliminating the unavoidable spurious resonances and a detailed discussion are provided in Ref.~\cite{pot12}. Thus we recall only the main ideas: for experimental data we showed that the HI should be applied several times with different sets of internal parameters, each giving a set of $\nu_j$ and $A_j$. Then the reconstruction based on these results is compared to the original signal. Figure~\ref{fig:resonanceClouds} shows part of a typical spectrum (black solid line) and the best (concerning the $\chi^2$ error) individual reconstruction (blue dashed line) within the window indicated by the vertical lines. The corresponding resonances are marked by blue triangles in the lower panel, the complex plane. The orange crosses belong to other resonance sets, also leading to good reconstructions (to maintain clarity they are not shown in the upper two panels), called good resonances. Other sets not meeting the criterion are rejected.

For the three-disk system we checked the reliability of the HI by comparing the experimental resonances with calculations based on the algorithm of Gaspard and Rice \cite{gas89d}. However even for experiments with closed microwave systems it is known that only the lowest resonances agree well with the zeta function predictions. For higher frequencies the experimental perturbations disturb the measured spectrum such that the measured resonances cannot be associated directly to the theoretical ones, but statistical properties such as the resonance density persist (see also Ref.~\onlinecite{fyo12}).

Figure~\ref{fig:WidthDistr} shows the good HI resonances in orange and the best in blue for $R/a = 5.5$, $ 40 \leq {\rm Re}\, k \leq 500\,{\rm m}^{-1}$. The orange poles form ``clouds'' around the blue triangles --the elongated shape of the clouds is a consequence of the nonisometric axis ranges. The black crosses indicate the numerically calculated resonances. The composition of resonance chains is typical for large $R/a$ parameters \cite{wir99b,lu03,gas89d}. The individual resonances are not reproduced by the experimental data due to inevitable reflections at the absorbers and the perturbation by the antenna but the resonance free regions and the resonance density coincide.

On the right of Fig.~\ref{fig:WidthDistr} the corresponding $\mathrm{Im}\,k$ distributions and $P(\mathrm{Im}\,k)$ are shown, in solid black for the numerically calculated and in dashed-dotted blue for the experimental spectrum. The distributions are the same within the limits of error. This was also true for all good reconstructions passing the $\chi^2$ criterion -- one example is shown in orange. In fact, one can show that agreement with $P(\mathrm{Im}\, k)$ is robust with respect to errors in the reconstruction as long as the number of resonances entering the reconstruction is approximately the same. For the example shown in Fig.~\ref{fig:WidthDistr} the number varied between 94 for the numerical data and 117 for the individual reconstruction.

By measuring the averaged $\mathrm{Im}\,k$ distribution for reduced three- and five-disk systems with different $R/a$, we can study the dependence of the $\mathrm{Im}\,k$ distribution on the opening of the system (see Fig.~\ref{fig:EscRate}). For varying $R/a$ the averaged histogram of $\mathrm{Im}\, k $ is plotted as a shade plot. The five-disk case is presented in Fig.~\ref{fig:EscRate}(a). For $R/a=2$ the system is completely closed; however we observe already a small gap $\approx 0.15\,$m$^{-1}$ due to antenna and wall absorbing effects. When opening the system the very narrow $\mathrm{Im}\, k $ distribution first gets wider and the maximum of the distribution moves towards higher imaginary parts. From $R/a\approx2.5$ the resonance free region starts to grow and reaches a value of $\approx 0.5\,$m$^{-1}$ for the maximal accessible opening at $R/a = 3.9$. Over the whole $R/a$ range the value of $P(1/2)$ stays positive, thus providing no lower bound on the spectral gap. The solid black line in the shade plot shows half the classical escape rate calculated by the cycle expansion. We show this curve only for $R/a>2.41$ as
for lower values, pruning starts for order 4 orbits, and the symbolic dynamic is no longer complete \cite{cvi89}. In agreement with high frequency calculations \cite{lu03}, our experiment in a much lower frequency regime shows that the maximum of the $\mathrm{Im}\,k$ distribution is described by $-\gamma_\mathrm{cl}/2$. We emphasize that there are no free parameters to fit $\gamma_\mathrm{cl}$ to the experiments.

The three-disk system is more open and the repeller becomes thin for $R/a \geq 2.83$; i.e., $P(1/2) < 0 $ provides a lower bound on the gap. Again one sees that the gap first increases and only for high $R/a$ values coincides with the lower bound $P(1/2)$ (dotted black line). At which exact value of $R/a$ the gap appears and whether it appears before $P(1/2)$ becomes negative is less clear than in the five-disk system. The area of the totally closed system ($R/a=2$) is too small to allow meaningful measurements. Since there are no pruned orbits until order 4 from $R/a = 2.01$ on, we are able to plot the calculated curve for the full measured range. The maximum of the $\mathrm{Im}\, k$ distribution decreases for $R/a>3$ which might be surprising at first sight. The reason is that the time of flight between two scattering events increases linearly which will overcompensate the defocusing effect of a scattering event for large enough $R/a$.

Figure~\ref{fig:EscRate}(c) shows the shade plot for the numerical data of the reduced three-disk system. Again the correspondence of $-\gamma_\mathrm{cl}/2$ (solid black line) is clearly visible only for larger $R/a$ values. For large $R/a$ the lower bound $P(1/2)$ (dotted black line) coincides well with the numerically observed gap. The first appearance of the gap is not described by $P(1/2)$ (see $2.5<R/a<2.83$). Here $P(1/2)$ is still positive but a clear gap is already visible; the same phenomenon as observed in the experimental data of the five-disk system. We note that $P(1/2)$ is a lower bound for the gap and that it is not optimal at high energies \cite{pet10,nau05}. It may happen in the experiment, but not in the hyperbolic quotient case, that some low energy resonances violate the semiclassical gap bound $P(1/2)$. However, we are restricted in the wave number range; thus, it is not guaranteed that we observe the optimal gap. For the numerical data [see Fig.~\ref{fig:EscRate}(c)] all imaginary parts calculated are below $P(1/2)$. The fact that there seem to be values below $P(1/2)$ is due to the bin size of the histogram. In the experimental spectra the small number of resonances within the gap correspond to spurious resonances which survived the filtering.

In this Letter we have demonstrated the existence of a spectral gap in open chaotic $n$-disk microwave systems. We could extract the resonances from the measured signal and thus had direct access to the gap and the maximum of the $\mathrm{Im}\, k $ distribution. These were compared with the calculated classical values for $P(1/2)$ and $- \gamma_\mathrm{cl}/2 = P ( 1 ) /2 $. A good agreement was found for sufficiently open systems. But we also show that the bound $P(1/2)$ does not describe the {opening of the gap} for experimental or numerical data. We would like to emphasize that all investigations were performed in the low lying $k$ regime thus showing a remarkable agreement with the semiclassical predictions.

We thank S. Nonnenmacher, B. Eckhardt for intensive discussions, S. M\"{o}ckel for providing C++ code, and DFG via the Forschergruppe 760, 'German National Academic Foundation' (T.W.), CNRS-INP via the program PEPSPTI (U. K.), and NSF via Grant No. DMS-1201417 (M. Z.) for partial support.


\begin{thebibliography}{41}%
\makeatletter
\providecommand \@ifxundefined [1]{%
 \@ifx{#1\undefined}
}%
\providecommand \@ifnum [1]{%
 \ifnum #1\expandafter \@firstoftwo
 \else \expandafter \@secondoftwo
 \fi
}%
\providecommand \@ifx [1]{%
 \ifx #1\expandafter \@firstoftwo
 \else \expandafter \@secondoftwo
 \fi
}%
\providecommand \natexlab [1]{#1}%
\providecommand \enquote  [1]{``#1''}%
\providecommand \bibnamefont  [1]{#1}%
\providecommand \bibfnamefont [1]{#1}%
\providecommand \citenamefont [1]{#1}%
\providecommand \href@noop [0]{\@secondoftwo}%
\providecommand \href [0]{\begingroup \@sanitize@url \@href}%
\providecommand \@href[1]{\@@startlink{#1}\@@href}%
\providecommand \@@href[1]{\endgroup#1\@@endlink}%
\providecommand \@sanitize@url [0]{\catcode `\\12\catcode `\$12\catcode
  `\&12\catcode `\#12\catcode `\^12\catcode `\_12\catcode `\%12\relax}%
\providecommand \@@startlink[1]{}%
\providecommand \@@endlink[0]{}%
\providecommand \url  [0]{\begingroup\@sanitize@url \@url }%
\providecommand \@url [1]{\endgroup\@href {#1}{\urlprefix }}%
\providecommand \urlprefix  [0]{URL }%
\providecommand \Eprint [0]{\href }%
\@ifxundefined \urlstyle {%
  \providecommand \doi  [0]{\begingroup \@sanitize@url \@doi}%
  \providecommand \@doi [1]{\endgroup \@@startlink {\doibase
  #1}doi:\discretionary {}{}{}#1\@@endlink }%
}{%
  \providecommand \doi  [0]{doi:\discretionary{}{}{}\begingroup
  \urlstyle{rm}\Url }%
}%
\providecommand \doibase [0]{http://dx.doi.org/}%
\providecommand \Doi [0]{\begingroup \@sanitize@url \@Doi }%
\providecommand \@Doi  [1]{\endgroup\@@startlink{\doibase#1}\@@Doi}%
\providecommand \@@Doi [1]{#1\@@endlink}%
\providecommand \selectlanguage [0]{\@gobble}%
\providecommand \bibinfo  [0]{\@secondoftwo}%
\providecommand \bibfield  [0]{\@secondoftwo}%
\providecommand \translation [1]{[#1]}%
\providecommand \BibitemOpen [0]{}%
\providecommand \bibitemStop [0]{}%
\providecommand \bibitemNoStop [0]{.\EOS\space}%
\providecommand \EOS [0]{\spacefactor3000\relax}%
\providecommand \BibitemShut  [1]{\csname bibitem#1\endcsname}%
\bibitem [{\citenamefont {Weyl}(1912)}]{wey12c}%
  \BibitemOpen
  \bibfield  {author} {\bibinfo {author} {\bibfnamefont {H.}~\bibnamefont
  {Weyl}},\ }\Doi {10.1007/BF01456804} {\bibfield  {journal} {\bibinfo
  {journal} {Math. Annalen},\ }\textbf {\bibinfo {volume} {71}},\ \bibinfo
  {pages} {441} (\bibinfo {year} {1912})}\BibitemShut {NoStop}%
\bibitem [{\citenamefont {Gutzwiller}(1971)}]{gut71}%
  \BibitemOpen
  \bibfield  {author} {\bibinfo {author} {\bibfnamefont {M.~C.}\ \bibnamefont
  {Gutzwiller}},\ }\href@noop {} {\bibfield  {journal} {\bibinfo  {journal} {J.
  Math. Phys.},\ }\textbf {\bibinfo {volume} {12}},\ \bibinfo {pages} {343}
  (\bibinfo {year} {1971})}\BibitemShut {NoStop}%
\bibitem [{\citenamefont {Gutzwiller}(1990)}]{gut90}%
  \BibitemOpen
  \bibfield  {author} {\bibinfo {author} {\bibfnamefont {M.~C.}\ \bibnamefont
  {Gutzwiller}},\ }\href@noop {} {\emph {\bibinfo {title} {Chaos in Classical
  and Quantum Mechanics}}},\ Interdisciplinary Applied Mathematics, Vol. 1\
  (\bibinfo  {publisher} {Springer},\ \bibinfo {address} {New York},\ \bibinfo
  {year} {1990})\BibitemShut {NoStop}%
\bibitem [{\citenamefont {Graefe}\ \emph {et~al.}(2010)\citenamefont {Graefe},
  \citenamefont {H{\"o}ning},\ and\ \citenamefont {Korsch}}]{gra10}%
  \BibitemOpen
  \bibfield  {author} {\bibinfo {author} {\bibfnamefont {E.-M.}\ \bibnamefont
  {Graefe}}, \bibinfo {author} {\bibfnamefont {M.}~\bibnamefont {H{\"o}ning}},
  \ and\ \bibinfo {author} {\bibfnamefont {H.~J.}\ \bibnamefont {Korsch}},\
  }\Doi {10.1088/1751-8113/43/7/075306} {\bibfield  {journal} {\bibinfo
  {journal} {J. Phys. A},\ }\textbf {\bibinfo {volume} {43}},\ \bibinfo {pages}
  {075306} (\bibinfo {year} {2010})}\BibitemShut {NoStop}%
\bibitem [{\citenamefont {Nonnenmacher}(2011)}]{non11}%
  \BibitemOpen
  \bibfield  {author} {\bibinfo {author} {\bibfnamefont {S.}~\bibnamefont
  {Nonnenmacher}},\ }\Doi {10.1088/0951-7715/24/12/R02} {\bibfield  {journal}
  {\bibinfo  {journal} {Nonlinearity},\ }\textbf {\bibinfo {volume} {24}},\
  \bibinfo {pages} {R123} (\bibinfo {year} {2011})}\BibitemShut {NoStop}%
\bibitem [{\citenamefont {Moiseyev}(1998)}]{moi98}%
  \BibitemOpen
  \bibfield  {author} {\bibinfo {author} {\bibfnamefont {N.}~\bibnamefont
  {Moiseyev}},\ }\Doi {10.1016/S0370-1573(98)00002-7} {\bibfield  {journal}
  {\bibinfo  {journal} {Phys. Rev.},\ }\textbf {\bibinfo {volume} {302}},\
  \bibinfo {pages} {212} (\bibinfo {year} {1998})},\ ISSN \bibinfo {issn}
  {0370-1573}\BibitemShut {NoStop}%
\bibitem [{\citenamefont {Zworski}(1999){\natexlab{a}}}]{zwo99a}%
  \BibitemOpen
  \bibfield  {author} {\bibinfo {author} {\bibfnamefont {M.}~\bibnamefont
  {Zworski}},\ }\href@noop {} {\bibfield  {journal} {\bibinfo  {journal} {Not.
  AMS},\ }\textbf {\bibinfo {volume} {43}},\ \bibinfo {pages} {319} (\bibinfo
  {year} {1999}{\natexlab{a}})}\BibitemShut {NoStop}%
\bibitem [{\citenamefont {Kuhl}\ \emph {et~al.}(2013)\citenamefont {Kuhl},
  \citenamefont {Legrand},\ and\ \citenamefont {Mortessagne}}]{kuh13}%
  \BibitemOpen
  \bibfield  {author} {\bibinfo {author} {\bibfnamefont {U.}~\bibnamefont
  {Kuhl}}, \bibinfo {author} {\bibfnamefont {O.}~\bibnamefont {Legrand}}, \
  and\ \bibinfo {author} {\bibfnamefont {F.}~\bibnamefont {Mortessagne}},\
  }\Doi {10.1002/prop.201200101} {\bibfield  {journal} {\bibinfo  {journal}
  {Fortschritte der Physik},\ }\textbf {\bibinfo {volume} {61}},\ \bibinfo
  {pages} {404} (\bibinfo {year} {2013})}\BibitemShut {NoStop}%
\bibitem [{\citenamefont {Sj{\"o}strand}(1990)}]{sjo90}%
  \BibitemOpen
  \bibfield  {author} {\bibinfo {author} {\bibfnamefont {J.}~\bibnamefont
  {Sj{\"o}strand}},\ }\Doi {10.1215/S0012-7094-90-06001-6} {\bibfield
  {journal} {\bibinfo  {journal} {Duke Math. J.},\ }\textbf {\bibinfo {volume}
  {60}},\ \bibinfo {pages} {1} (\bibinfo {year} {1990})}\BibitemShut {NoStop}%
\bibitem [{\citenamefont {Sj{\"o}strand}\ and\ \citenamefont
  {Zworski}(2007)}]{sjo07}%
  \BibitemOpen
  \bibfield  {author} {\bibinfo {author} {\bibfnamefont {J.}~\bibnamefont
  {Sj{\"o}strand}}\ and\ \bibinfo {author} {\bibfnamefont {M.}~\bibnamefont
  {Zworski}},\ }\Doi {10.1215/S0012-7094-07-13731-1} {\bibfield  {journal}
  {\bibinfo  {journal} {Duke Math. J.},\ }\textbf {\bibinfo {volume} {137}},\
  \bibinfo {pages} {381} (\bibinfo {year} {2007})}\BibitemShut {NoStop}%
\bibitem [{\citenamefont {Schomerus}\ \emph {et~al.}(2000)\citenamefont
  {Schomerus}, \citenamefont {Frahm}, \citenamefont {Patra},\ and\
  \citenamefont {Beenakker}}]{sch00d}%
  \BibitemOpen
  \bibfield  {author} {\bibinfo {author} {\bibfnamefont {H.}~\bibnamefont
  {Schomerus}}, \bibinfo {author} {\bibfnamefont {K.~M.}\ \bibnamefont
  {Frahm}}, \bibinfo {author} {\bibfnamefont {M.}~\bibnamefont {Patra}}, \ and\
  \bibinfo {author} {\bibfnamefont {C.~W.~J.}\ \bibnamefont {Beenakker}},\
  }\href@noop {} {\bibfield  {journal} {\bibinfo  {journal} {Physica A},\
  }\textbf {\bibinfo {volume} {278}},\ \bibinfo {pages} {469} (\bibinfo {year}
  {2000})}\BibitemShut {NoStop}%
\bibitem [{\citenamefont {Lin}(2002)}]{lin02b}%
  \BibitemOpen
  \bibfield  {author} {\bibinfo {author} {\bibfnamefont {K.~K.}\ \bibnamefont
  {Lin}},\ }\Doi {10.1006/jcph.2001.6986} {\bibfield  {journal} {\bibinfo
  {journal} {J. Comp. Phys.},\ }\textbf {\bibinfo {volume} {176}},\ \bibinfo
  {pages} {295} (\bibinfo {year} {2002})}\BibitemShut {NoStop}%
\bibitem [{\citenamefont {Lu}\ \emph {et~al.}(2003)\citenamefont {Lu},
  \citenamefont {Sridhar},\ and\ \citenamefont {Zworski}}]{lu03}%
  \BibitemOpen
  \bibfield  {author} {\bibinfo {author} {\bibfnamefont {W.~T.}\ \bibnamefont
  {Lu}}, \bibinfo {author} {\bibfnamefont {S.}~\bibnamefont {Sridhar}}, \ and\
  \bibinfo {author} {\bibfnamefont {M.}~\bibnamefont {Zworski}},\ }\href@noop
  {} {\bibfield  {journal} {\bibinfo  {journal} {Phys. Rev. Lett.},\ }\textbf
  {\bibinfo {volume} {91}},\ \bibinfo {pages} {154101} (\bibinfo {year}
  {2003})}\BibitemShut {NoStop}%
\bibitem [{\citenamefont {Schomerus}\ \emph {et~al.}(2009)\citenamefont
  {Schomerus}, \citenamefont {Wiersig},\ and\ \citenamefont {Main}}]{sch09}%
  \BibitemOpen
  \bibfield  {author} {\bibinfo {author} {\bibfnamefont {H.}~\bibnamefont
  {Schomerus}}, \bibinfo {author} {\bibfnamefont {J.}~\bibnamefont {Wiersig}},
  \ and\ \bibinfo {author} {\bibfnamefont {J.}~\bibnamefont {Main}},\ }\Doi
  {10.1103/PhysRevA.79.053806} {\bibfield  {journal} {\bibinfo  {journal}
  {Phys. Rev. A},\ }\textbf {\bibinfo {volume} {79}},\ \bibinfo {pages}
  {053806} (\bibinfo {year} {2009})}\BibitemShut {NoStop}%
\bibitem [{\citenamefont {Potzuweit}\ \emph {et~al.}(2012)\citenamefont
  {Potzuweit}, \citenamefont {Weich}, \citenamefont {Barkhofen}, \citenamefont
  {Kuhl}, \citenamefont {St{\"o}ckmann},\ and\ \citenamefont
  {Zworski}}]{pot12}%
  \BibitemOpen
  \bibfield  {author} {\bibinfo {author} {\bibfnamefont {A.}~\bibnamefont
  {Potzuweit}}, \bibinfo {author} {\bibfnamefont {T.}~\bibnamefont {Weich}},
  \bibinfo {author} {\bibfnamefont {S.}~\bibnamefont {Barkhofen}}, \bibinfo
  {author} {\bibfnamefont {U.}~\bibnamefont {Kuhl}}, \bibinfo {author}
  {\bibfnamefont {H.-J.}\ \bibnamefont {St{\"o}ckmann}}, \ and\ \bibinfo
  {author} {\bibfnamefont {M.}~\bibnamefont {Zworski}},\ }\Doi
  {10.1103/PhysRevE.86.066205} {\bibfield  {journal} {\bibinfo  {journal}
  {Phys. Rev. E},\ }\textbf {\bibinfo {volume} {86}},\ \bibinfo {pages}
  {066205} (\bibinfo {year} {2012})}\BibitemShut {NoStop}%
\bibitem [{\citenamefont {Ikawa}(1988)}]{ika88}%
  \BibitemOpen
  \bibfield  {author} {\bibinfo {author} {\bibfnamefont {M.}~\bibnamefont
  {Ikawa}},\ }\Doi {10.5802/aif.1137} {\bibfield  {journal} {\bibinfo
  {journal} {Ann. Inst. Fourier},\ }\textbf {\bibinfo {volume} {38}},\ \bibinfo
  {pages} {113} (\bibinfo {year} {1988})}\BibitemShut {NoStop}%
\bibitem [{\citenamefont {Gaspard}\ and\ \citenamefont
  {Rice}(1989){\natexlab{a}}}]{gas89a}%
  \BibitemOpen
  \bibfield  {author} {\bibinfo {author} {\bibfnamefont {P.}~\bibnamefont
  {Gaspard}}\ and\ \bibinfo {author} {\bibfnamefont {S.~A.}\ \bibnamefont
  {Rice}},\ }\Doi {http://dx.doi.org/10.1063/1.456018} {\bibfield  {journal}
  {\bibinfo  {journal} {J. Chem. Phys.},\ }\textbf {\bibinfo {volume} {90}},\
  \bibinfo {pages} {2242} (\bibinfo {year} {1989}{\natexlab{a}})}\BibitemShut
  {NoStop}%
\bibitem [{\citenamefont {Gaspard}\ and\ \citenamefont
  {Rice}(1989){\natexlab{b}}}]{gas89b}%
  \BibitemOpen
  \bibfield  {author} {\bibinfo {author} {\bibfnamefont {P.}~\bibnamefont
  {Gaspard}}\ and\ \bibinfo {author} {\bibfnamefont {S.~A.}\ \bibnamefont
  {Rice}},\ }\Doi {10.1063/1.456017} {\bibfield  {journal} {\bibinfo  {journal}
  {J. Chem. Phys.},\ }\textbf {\bibinfo {volume} {90}},\ \bibinfo {pages}
  {2225} (\bibinfo {year} {1989}{\natexlab{b}})}\BibitemShut {NoStop}%
\bibitem [{\citenamefont {Gaspard}\ and\ \citenamefont
  {Rice}(1989){\natexlab{c}}}]{gas89d}%
  \BibitemOpen
  \bibfield  {author} {\bibinfo {author} {\bibfnamefont {P.}~\bibnamefont
  {Gaspard}}\ and\ \bibinfo {author} {\bibfnamefont {S.~A.}\ \bibnamefont
  {Rice}},\ }\Doi {10.1063/1.456019} {\bibfield  {journal} {\bibinfo  {journal}
  {J. Chem. Phys.},\ }\textbf {\bibinfo {volume} {90}},\ \bibinfo {pages}
  {2255} (\bibinfo {year} {1989}{\natexlab{c}})}\BibitemShut {NoStop}%
\bibitem [{\citenamefont {Cvitanovi\'c}\ and\ \citenamefont
  {Eckhardt}(1989)}]{cvi89}%
  \BibitemOpen
  \bibfield  {author} {\bibinfo {author} {\bibfnamefont {P.}~\bibnamefont
  {Cvitanovi\'c}}\ and\ \bibinfo {author} {\bibfnamefont {B.}~\bibnamefont
  {Eckhardt}},\ }\href@noop {} {\bibfield  {journal} {\bibinfo  {journal}
  {Phys. Rev. Lett.},\ }\textbf {\bibinfo {volume} {63}},\ \bibinfo {pages}
  {823} (\bibinfo {year} {1989})}\BibitemShut {NoStop}%
\bibitem [{\citenamefont {Eckhardt}\ \emph {et~al.}(1995)\citenamefont
  {Eckhardt}, \citenamefont {Russberg}, \citenamefont {Cvitanovi\'{c}},
  \citenamefont {Rosenqvist},\ and\ \citenamefont {Scherer}}]{eck95b}%
  \BibitemOpen
  \bibfield  {author} {\bibinfo {author} {\bibfnamefont {B.}~\bibnamefont
  {Eckhardt}}, \bibinfo {author} {\bibfnamefont {G.}~\bibnamefont {Russberg}},
  \bibinfo {author} {\bibfnamefont {P.}~\bibnamefont {Cvitanovi\'{c}}},
  \bibinfo {author} {\bibfnamefont {P.}~\bibnamefont {Rosenqvist}}, \ and\
  \bibinfo {author} {\bibfnamefont {P.}~\bibnamefont {Scherer}},\ }in\
  \href@noop {} {\emph {\bibinfo {booktitle} {Quantum Chaos Between Order and
  Disorder}}},\ \bibinfo {editor} {edited by\ \bibinfo {editor} {\bibfnamefont
  {C.}~\bibnamefont {Casati}}\ and\ \bibinfo {editor} {\bibfnamefont
  {B.}~\bibnamefont {Chirikov}}}\ (\bibinfo  {publisher} {University Press},\
  \bibinfo {address} {Cambridge},\ \bibinfo {year} {1995})\ p.\ \bibinfo
  {pages} {405}\BibitemShut {NoStop}%
\bibitem [{\citenamefont {Bowen}(1979)}]{bow79}%
  \BibitemOpen
  \bibfield  {author} {\bibinfo {author} {\bibfnamefont {R.}~\bibnamefont
  {Bowen}},\ }\Doi {10.1007/BF02684767} {\bibfield  {journal} {\bibinfo
  {journal} {Inst. Hautes {\'E}tudes Sci. Publ. Math.},\ }\textbf {\bibinfo
  {volume} {50}},\ \bibinfo {pages} {11} (\bibinfo {year} {1979})}\BibitemShut
  {NoStop}%
\bibitem [{\citenamefont {Nonnenmacher}\ and\ \citenamefont
  {Zworski}(2009)}]{non09}%
  \BibitemOpen
  \bibfield  {author} {\bibinfo {author} {\bibfnamefont {S.}~\bibnamefont
  {Nonnenmacher}}\ and\ \bibinfo {author} {\bibfnamefont {M.}~\bibnamefont
  {Zworski}},\ }\Doi {10.1007/s11511-009-0041-z} {\bibfield  {journal}
  {\bibinfo  {journal} {Acta Mathematica},\ }\textbf {\bibinfo {volume}
  {203}},\ \bibinfo {pages} {149} (\bibinfo {year} {2009})}\BibitemShut
  {NoStop}%
\bibitem [{\citenamefont {Petkov}\ and\ \citenamefont
  {Stoyanov}(2010)}]{pet10}%
  \BibitemOpen
  \bibfield  {author} {\bibinfo {author} {\bibfnamefont {V.}~\bibnamefont
  {Petkov}}\ and\ \bibinfo {author} {\bibfnamefont {L.}~\bibnamefont
  {Stoyanov}},\ }\href@noop {} {\bibfield  {journal} {\bibinfo  {journal}
  {Anal. PDE},\ }\textbf {\bibinfo {volume} {3}},\ \bibinfo {pages} {427}
  (\bibinfo {year} {2010})}\BibitemShut {NoStop}%
\bibitem [{\citenamefont {Patterson}(1976)}]{pat76}%
  \BibitemOpen
  \bibfield  {author} {\bibinfo {author} {\bibfnamefont {S.~J.}\ \bibnamefont
  {Patterson}},\ }\Doi {10.1007/BF02392046} {\bibfield  {journal} {\bibinfo
  {journal} {Acta Mathematica},\ }\textbf {\bibinfo {volume} {136}},\ \bibinfo
  {pages} {241} (\bibinfo {year} {1976})}\BibitemShut {NoStop}%
\bibitem [{\citenamefont {Sullivan}(1979)}]{sul79}%
  \BibitemOpen
  \bibfield  {author} {\bibinfo {author} {\bibfnamefont {D.}~\bibnamefont
  {Sullivan}},\ }\Doi {10.1007/BF02684773} {\bibfield  {journal} {\bibinfo
  {journal} {Inst. Hautes {\'E}tudes Sci. Publ. Math.},\ }\textbf {\bibinfo
  {volume} {50}},\ \bibinfo {pages} {171} (\bibinfo {year} {1979})}\BibitemShut
  {NoStop}%
\bibitem [{\citenamefont {Borthwick}(2007)}]{bor07}%
  \BibitemOpen
  \bibfield  {author} {\bibinfo {author} {\bibfnamefont {D.}~\bibnamefont
  {Borthwick}},\ }\href@noop {} {\emph {\bibinfo {title} {Spectral Theory of
  Infinite-Area Hyperbolic Surfaces}}},\ \bibinfo {series} {Progress in
  Mathematics}, Vol.\ \bibinfo {volume} {256}\ (\bibinfo  {publisher}
  {Birkh{\"a}user},\ \bibinfo {address} {Boston},\ \bibinfo {year}
  {2007})\BibitemShut {NoStop}%
\bibitem [{\citenamefont {Naud}(2005)}]{nau05}%
  \BibitemOpen
  \bibfield  {author} {\bibinfo {author} {\bibfnamefont {F.}~\bibnamefont
  {Naud}},\ }\Doi {10.1016/j.ansens.2004.11.002} {\bibfield  {journal}
  {\bibinfo  {journal} {Ann. Sci. Ecole Norm. Sup.},\ }\textbf {\bibinfo
  {volume} {38}},\ \bibinfo {pages} {116} (\bibinfo {year} {2005})}\BibitemShut
  {NoStop}%
\bibitem [{\citenamefont {Bourgain}\ \emph {et~al.}(2011)\citenamefont
  {Bourgain}, \citenamefont {Gamburd},\ and\ \citenamefont {Sarnak.}}]{bou11}%
  \BibitemOpen
  \bibfield  {author} {\bibinfo {author} {\bibfnamefont {J.}~\bibnamefont
  {Bourgain}}, \bibinfo {author} {\bibfnamefont {A.}~\bibnamefont {Gamburd}}, \
  and\ \bibinfo {author} {\bibfnamefont {P.}~\bibnamefont {Sarnak.}},\
  }\href@noop {} {\bibfield  {journal} {\bibinfo  {journal} {Acta
  Mathematica},\ }\textbf {\bibinfo {volume} {207}},\ \bibinfo {pages} {255}
  (\bibinfo {year} {2011})}\BibitemShut {NoStop}%
\bibitem [{\citenamefont {Naud}(2013)}]{nau13}%
  \BibitemOpen
  \bibfield  {author} {\bibinfo {author} {\bibfnamefont {F.}~\bibnamefont
  {Naud}},\ }\Doi {10.1007/s00222-013-0463-2} {\bibfield  {journal} {\bibinfo
  {journal} {Inventiones mathematicae},\ }\textbf {\bibinfo {volume} {March}},\
  \bibinfo {pages} {1} (\bibinfo {year} {2013})}\BibitemShut {NoStop}%
\bibitem [{\citenamefont {Guillop{\'e}}\ \emph {et~al.}(2004)\citenamefont
  {Guillop{\'e}}, \citenamefont {Lin},\ and\ \citenamefont {Zworski}}]{gui04}%
  \BibitemOpen
  \bibfield  {author} {\bibinfo {author} {\bibfnamefont {L.}~\bibnamefont
  {Guillop{\'e}}}, \bibinfo {author} {\bibfnamefont {K.~K.}\ \bibnamefont
  {Lin}}, \ and\ \bibinfo {author} {\bibfnamefont {M.}~\bibnamefont
  {Zworski}},\ }\Doi {10.1007/s00220-003-1007-1} {\bibfield  {journal}
  {\bibinfo  {journal} {Commun. Math. Phys.},\ }\textbf {\bibinfo {volume}
  {245}},\ \bibinfo {pages} {149} (\bibinfo {year} {2004})}\BibitemShut
  {NoStop}%
\bibitem [{\citenamefont {Zworski}(1999){\natexlab{b}}}]{zwo99b}%
  \BibitemOpen
  \bibfield  {author} {\bibinfo {author} {\bibfnamefont {M.}~\bibnamefont
  {Zworski}},\ }\Doi {10.1007/s002220050313} {\bibfield  {journal} {\bibinfo
  {journal} {Inventiones mathematicae},\ }\textbf {\bibinfo {volume} {136}},\
  \bibinfo {pages} {353} (\bibinfo {year} {1999}{\natexlab{b}})}\BibitemShut
  {NoStop}%
\bibitem [{\citenamefont {Datchev}\ and\ \citenamefont
  {Dyatlov}(2012)}]{arXdat12}%
  \BibitemOpen
  \bibfield  {author} {\bibinfo {author} {\bibfnamefont {K.}~\bibnamefont
  {Datchev}}\ and\ \bibinfo {author} {\bibfnamefont {S.}~\bibnamefont
  {Dyatlov}},\ }\href@noop {} {\enquote {\bibinfo {title} {Fractal weyl laws
  for asymptotically hyperbolic manifolds},}\ }\bibinfo {howpublished}
  {Preprint} (\bibinfo {year} {2012}),\ \bibinfo {note}
  {arXiv:1206.2255v3}\BibitemShut {NoStop}%
\bibitem [{\citenamefont {Lu}\ \emph {et~al.}(1999)\citenamefont {Lu},
  \citenamefont {Rose}, \citenamefont {Pance},\ and\ \citenamefont
  {Sridhar}}]{lu99}%
  \BibitemOpen
  \bibfield  {author} {\bibinfo {author} {\bibfnamefont {W.}~\bibnamefont
  {Lu}}, \bibinfo {author} {\bibfnamefont {M.}~\bibnamefont {Rose}}, \bibinfo
  {author} {\bibfnamefont {K.}~\bibnamefont {Pance}}, \ and\ \bibinfo {author}
  {\bibfnamefont {S.}~\bibnamefont {Sridhar}},\ }\href@noop {} {\bibfield
  {journal} {\bibinfo  {journal} {Phys. Rev. Lett.},\ }\textbf {\bibinfo
  {volume} {82}},\ \bibinfo {pages} {5233} (\bibinfo {year}
  {1999})}\BibitemShut {NoStop}%
\bibitem [{\citenamefont {Lu}\ \emph {et~al.}(2000)\citenamefont {Lu},
  \citenamefont {Viola}, \citenamefont {Pance}, \citenamefont {Rose},\ and\
  \citenamefont {Sridhar}}]{lu00}%
  \BibitemOpen
  \bibfield  {author} {\bibinfo {author} {\bibfnamefont {W.}~\bibnamefont
  {Lu}}, \bibinfo {author} {\bibfnamefont {L.}~\bibnamefont {Viola}}, \bibinfo
  {author} {\bibfnamefont {K.}~\bibnamefont {Pance}}, \bibinfo {author}
  {\bibfnamefont {M.}~\bibnamefont {Rose}}, \ and\ \bibinfo {author}
  {\bibfnamefont {S.}~\bibnamefont {Sridhar}},\ }\href@noop {} {\bibfield
  {journal} {\bibinfo  {journal} {Phys. Rev. E},\ }\textbf {\bibinfo {volume}
  {61}},\ \bibinfo {pages} {3652} (\bibinfo {year} {2000})}\BibitemShut
  {NoStop}%
\bibitem [{\citenamefont {St{\"o}ckmann}(1999)}]{stoe99}%
  \BibitemOpen
  \bibfield  {author} {\bibinfo {author} {\bibfnamefont {H.-J.}\ \bibnamefont
  {St{\"o}ckmann}},\ }\href@noop {} {\emph {\bibinfo {title} {Quantum Chaos -
  An Introduction}}}\ (\bibinfo  {publisher} {University Press},\ \bibinfo
  {address} {Cambridge},\ \bibinfo {year} {1999})\BibitemShut {NoStop}%
\bibitem [{\citenamefont {Stein}\ \emph {et~al.}(1995)\citenamefont {Stein},
  \citenamefont {St{\"o}ckmann},\ and\ \citenamefont {Stoffregen}}]{ste95}%
  \BibitemOpen
  \bibfield  {author} {\bibinfo {author} {\bibfnamefont {J.}~\bibnamefont
  {Stein}}, \bibinfo {author} {\bibfnamefont {H.-J.}\ \bibnamefont
  {St{\"o}ckmann}}, \ and\ \bibinfo {author} {\bibfnamefont {U.}~\bibnamefont
  {Stoffregen}},\ }\Doi {10.1103/PhysRevLett.75.53} {\bibfield  {journal}
  {\bibinfo  {journal} {Phys. Rev. Lett.},\ }\textbf {\bibinfo {volume} {75}},\
  \bibinfo {pages} {53} (\bibinfo {year} {1995})}\BibitemShut {NoStop}%
\bibitem [{\citenamefont {Main}(1999)}]{mai99}%
  \BibitemOpen
  \bibfield  {author} {\bibinfo {author} {\bibfnamefont {J.}~\bibnamefont
  {Main}},\ }\href@noop {} {\bibfield  {journal} {\bibinfo  {journal} {Phys.
  Rep.},\ }\textbf {\bibinfo {volume} {316}},\ \bibinfo {pages} {233} (\bibinfo
  {year} {1999})}\BibitemShut {NoStop}%
\bibitem [{\citenamefont {Kuhl}\ \emph {et~al.}(2008)\citenamefont {Kuhl},
  \citenamefont {H{\"o}hmann}, \citenamefont {Main},\ and\ \citenamefont
  {St{\"o}ckmann}}]{kuh08b}%
  \BibitemOpen
  \bibfield  {author} {\bibinfo {author} {\bibfnamefont {U.}~\bibnamefont
  {Kuhl}}, \bibinfo {author} {\bibfnamefont {R.}~\bibnamefont {H{\"o}hmann}},
  \bibinfo {author} {\bibfnamefont {J.}~\bibnamefont {Main}}, \ and\ \bibinfo
  {author} {\bibfnamefont {H.-J.}\ \bibnamefont {St{\"o}ckmann}},\ }\Doi
  {10.1103/PhysRevLett.100.254101} {\bibfield  {journal} {\bibinfo  {journal}
  {Phys. Rev. Lett.},\ }\textbf {\bibinfo {volume} {100}},\ \bibinfo {pages}
  {254101} (\bibinfo {year} {2008})}\BibitemShut {NoStop}%
\bibitem [{\citenamefont {Fyodorov}\ and\ \citenamefont {Savin}(2012)}]{fyo12}%
  \BibitemOpen
  \bibfield  {author} {\bibinfo {author} {\bibfnamefont {Y.~V.}\ \bibnamefont
  {Fyodorov}}\ and\ \bibinfo {author} {\bibfnamefont {D.~V.}\ \bibnamefont
  {Savin}},\ }\Doi {10.1103/PhysRevLett.108.184101} {\bibfield  {journal}
  {\bibinfo  {journal} {Phys. Rev. Lett.},\ }\textbf {\bibinfo {volume}
  {108}},\ \bibinfo {pages} {184101} (\bibinfo {year} {2012})}\BibitemShut
  {NoStop}%
\bibitem [{\citenamefont {Wirzba}(1999)}]{wir99b}%
  \BibitemOpen
  \bibfield  {author} {\bibinfo {author} {\bibfnamefont {A.}~\bibnamefont
  {Wirzba}},\ }\href@noop {} {\bibfield  {journal} {\bibinfo  {journal} {Phys.
  Rep.},\ }\textbf {\bibinfo {volume} {309}},\ \bibinfo {pages} {1} (\bibinfo
  {year} {1999})}\BibitemShut {NoStop}%
\end{thebibliography}
\end{document}